# Rotation Diffusion as an Additional Mechanism of Energy Dissipation in Polymer Melts

A.N.Yakunin[*]

*Karpov Institute of Physical Chemistry, 105064 Moscow, Russia*

**Abstract:** The research is important for a molecular theory of liquid and has a wide interest as an example solving the problem when dynamic parameters of systems can be indirectly connected with their equilibrium properties. In frameworks of the reptation model the power law with the 3.4-exponent for the melt viscosity relation to the molecular weight of linear flexible-chain polymer is predicted as distinct from the value 3 expected for a melt of ring macromolecules. To find the exponent close to experimental values it should be taken into account the rotation vibration precession motion of chain ends about the polymer melt flow direction.

---

[*] E-mail: yakunin@cc.nifhi.ac.ru

# 1 Introduction

In present, one may describe the polymer melt viscosity dependence $\eta(N)$ on the polymerisation degree $N$ (the number of chain monomers per one macromolecule) in terms of the reptation model [1-3] suggested by P.-G. de Gennes in 1971. The reptation theory explains why the power law for the melt viscosity relation to the molecular weight of polymer can be observed. However, experimental data show the strong discrepancy with the theory (the exponent is equal to 3.3-3.4 or even is out of the range as distinct from the theoretical value which is equal to 3). Usually, in order to achieve better agreement with experiment one may assume that the length of tube, created by entanglement chains and along that a macromolecule can crawl making the reptation motions, can be subjected to fluctuations [2]. Wool [4] has also obtained an analogous result using a different concept. Other attempts have been aimed modifying dynamic equations [5, 6]. They continue up to now. More complex systems such as phase-separate polymer solutions, blends, block and graft copolymer mesophases and other fluids are suppose to be applicable to studying by similar techniques [7]. However, in such approaches the number of monomers between neighbour entanglements along the chain, $N_e$, is introduced phenomenologically or ignored completely. Moreover, the recent Monte Carlo results [8] have shown that reptation motions prevail in melts where entropic trapping is absent in contrast to swollen gels [9]. At last, there exist models in which an exponential increase of the reptation time with increasing the polymerisation degree can be observed if polymer chains are long enough $N > N_e^3$ [10, 11 and refs ibidem]. Thus, the questions dealing with polymer melt viscosity and entanglements have been extensively discussed during the past decade [7, 8, 10-12]. The fact claims that the problems are open in the present time although the entanglement concentration has been already estimated (see [4, 13] for example).

The aim of the article is to construct the solution by such way which could enable to achieve a good accordance with the well-known experimental data in frameworks of the classical reptation model [1] without the use of any additional simplifying assumptions.

The structure of the article is following. We will recall the main principles of the reptation theory [1-3] and try to understand which of ways can be fitted to modify the theory. We will see that the model [14] based on fluctuations of tube length is mistaken. It should be introduced a mechanical field. In that case, a chain seems can "swell" in a medium of other chains and the vector connecting the chain ends can rotate (Figure 1) making a vibration precession motion about the direction of polymer melt flow. Finally, we will obtain a good agreement with the experimental data by help of the suggested changed reptation theory. Physical transparency of concepts forming its background makes very attractive this approach in understanding and explaining the nature of energy dissipation in polymer liquids.

## 2 Results and discussion

### 2.1 Condition for the reptation motion

Let us consider the reptation theory [1, 15]. A macromolecule of flexible-chain polymer melt crawls along the tube under the influence of an applied constant force, $f$, with the resulting velocity, $V$. Let $S_0$ be the cross-section area of tube. Assuming $S_0^{1/2} = aN_e^{1/2}$ where $a$ is the diameter of monomer (diameter of chain or the lattice constant, and we have defined $N_e$ above) one can find the expression for the length of tube, $L$ (in units of tube diameter):

$$L = S_0^{1/2}N/N_e = aN/N_e^{1/2}. \tag{1}$$

Here, $N$ is the polymerisation degree. It should be underlined that we are not going to write numerical constants in our formulae because we only want to obtain the true exponent for the melt viscosity dependence on the molecular weight of polymer.

The mobility of the chain along the tube, $\mu_t = 1/\zeta_{fr}$, may be defined as

$$\mu_t = V/f \tag{2}$$

where $\zeta_{fr}$ is the friction coefficient. Finally, using (1) and (2) we can find the time of relaxation (so-called the maximum time of relaxation), $\tau$:

$$\tau = L^2/D_t = (aN/N_e^{1/2})^2/(VTN^{-1}/f_1) = \tau_1 N^3/N_e \tag{3}$$

where $D_t = \mu_t T = VTN^{-1}/f_1$ is the diffusion coefficient of chain along the tube, $f_1$ is the friction force per one monomer, $T$ is the temperature expressed in energy units, $\tau_1 = a^2/D_1$ and $D_1$ are the time of relaxation and the diffusion coefficient typical for low-molecular-weight liquids, respectively. Thus, from (3) the main result of the reptation theory may be obtained: the polymer melt viscosity

$$\eta \sim E\tau \tag{4}$$

is proportional to the polymerisation degree to the third power; here,

$$E \sim Ta^{-3}/N_e \tag{5}$$

is the elastic modulus of fluctuation network of polymer melt [15].

In order to observe the reptation motions one can suppose that the following inequality should be held:

$$f_1 \, aN/T \ll 1, \tag{6}$$

i. e. the work against the friction force is much less than the thermal energy. It is true for flexible enough chains. Their conformations may be changed easy, and such chains will make these motions. Other assumptions lead to polymer glass with frozen conformations. Multiplying the both hands of this inequality by $aN$, we find

$$\tau_R V \ll aN \qquad (7)$$

where

$$\tau_R \sim \zeta_1 a^2 N^2 / T \qquad (8)$$

is the relaxation time of the first Rouse mode, $\zeta_1 = T/D_1$ is the friction coefficient of one monomer; another form of the inequality (7) can be written as follows:

$$\xi^2 \gg r^2 \qquad (9)$$

where $\xi^2 = Ta/f_1$ and $r \sim aN^{1/2}$ is the mean end-to-end distance.

It is the most important and interesting result of the present work. In reptational dynamics a characteristic spatial scale appears. Although it is connected with the friction force per one monomer, it is much more than the mean size of polymer coil.

Apart from "dynamic" properties of polymer melts the conditions (6, 7, 9) determine also their "static" properties or the lifetime of entanglement network (see (5)): if $N < N_e$ then since $\tau \sim \tau_1 N^2$ (one can regard as $N_e \sim N$ in (3) and $r \sim aN$ in (9)) we see that

$$f_1 a N_e^2 \sim T \qquad (10)$$

and $\eta \sim Ta^{-3} \tau_1 N$ from (3)-(5), i. e. another regime of viscous flow realizes in the absence of entanglement fluctuation network (here, formula (5) transforms to $E \sim Ta^{-3}/N$). In other words, $N_e$ is a thermodynamic parameter of polymer melts [13]. Note that the authors of work [10] call the value of $T/N_e^2$ as an effective difference of chemical potentials. Their results can not be compared with our estimations (6, 7, 9) since they have studied rather a polymer glass then a melt in accordance with their model, i. e. the authors are beyond frameworks of the classical reptation theory [1, 15].

2.2 Screening the volume interactions in an external field

In this work we suppose that viscous flow weakens the screening monomer-monomer interactions in a polymer melt which leads to enhanced effective attraction between monomers. Both correlation and screening (i. e. response) properties of a polymer melt in a steady state in general differ from those in equilibrium. This is indeed a special problem insufficiently discussed in literature, unfortunately. Let us obtain our estimations more rigorously in terms of pair correlation functions.

The energy, $F$, of interaction of two probe particles having the same chemical potential (for simplicity) as monomers of chain in a concentrated solution and situated at points $r_1$ and $r_2$ can be written as follows [3]

$$F(r_1 - r_2) = u(r_1 - r_2) + \varphi(r_1 - r_2)$$

where $u(r_1 - r_2)$ is the potential of interaction of these probe particles with each other and

$$\varphi(r_1 - r_2) = -(3vT/(\pi a^2 |r_1 - r_2|)) \exp(-|r_1 - r_2|/\xi_E) \tag{11}$$

can be received by standard techniques [3]. Here, the potential $\varphi(r_1 - r_2)$ is proportional to the Ornstein – Zernike correlation function [15, 16],

$$\xi_E = (a^2/(12vc))^{1/2} \tag{12}$$

is the Edwards correlation radius [15, 16], $c$ is the number of chain monomers per unit volume, $v$ is the parameter of excluded volume [15, 16]. According to the Flory theorem obtained in 1949, screening the volume interactions is a physical property of polymer melts [3, 15]

$$v_{eff} \sim v/N \tag{13}$$

where $v_{eff}$ is the effective parameter of excluded volume in polymer melts. From (12) one can obtain that $\xi_E \sim a$ since $vc=1$. $r \sim aN^{1/2}$ is the mean end-to-end distance of chain in melts [3, 15]. However, we have seen from (9) that there exists a spatial $\xi >> r$. This gives an opportunity to assume that if the reptation motions due to the action of a mechanical field in polymer melts take place indeed then i) screening the volume interactions vanishes ($v_{eff} \sim v$ in (13)), ii) the fluctuation attraction of chain ends and other monomers remains (11), iii) the correlation radius can increase.

Note that although all motions associated with the chain ends are fast at small scales of order of several atom radii, an inhibition of such motions of ends at large scales (of order of end-to-end distance) reduces the overall chain mobility.

2.3 Translation diffusion

Another mechanism of motions associated with the chain ends has been considered by M. Doi [14]. He has suggested that the mean-square displacement of ends along the tube is

$$<\xi^2(t)> = a^2 N (t/\tau_R)^{1/2} = (D_l t\, a^2)^{1/2} \qquad (14)$$

(see (8)) for $t \leq a^2 N/D_l \sim \tau_R$, that is, on the short time scale the chain end moves around quickly within the distance $aN^{1/2}$. However, this formulation contradicts our conclusion concerning the fluctuation attraction of chain ends. It should be discussed this question more accurately.

Let $g$ be the number of Rouse segments belonging to a small part of macromolecule. If $t \leq \tau_R = a^2 g^2/D_l$ then $t$ is the maximum relaxation time for the given part of chain [3]. Thus,

$$g = (D_l t/a^2)^{1/2}.$$

The mean-square displacement of the centre of gravity for this part of macromolecule can be written as follows [3]

$$D_l t/g \sim (D_l t\, a^2)^{1/2}. \qquad (15)$$

The equation (15) corresponds to the expression (14). But on the maximum relaxation time scale the mean-square displacement of the centre of gravity for the coil should be equal approximately to the mean-square displacement for a separate Rouse segment which can be expressed by the same formula (15). The important conclusion follows this reasoning: no translational motions give additional contributions in viscosity. It should be taken into account rotation diffusion motions in order to use the formula (1) with no changes. Note that we follow the authors of textbook [3] literally but result in other conclusions. We have used their arguments in our purposes.

2.4 Rotation diffusion

Let us consider the mechanical field in which the frequency, $\omega$, of rotation vibration motion of chain about the polymer melt flow direction is much less than the reciprocal value of $\tau$

$$\varphi = \omega\tau << 1.$$

Then the arising phase difference, $\varphi$, can be assumed [17] to be connected with a resulting torque which tends to return the chain ends in their initial position (see Figure 1). The scaling expression for the coefficient of rotation diffusion $D_{rot} \sim \tau^{-1}$ can be written as follows

$$D_{rot} \sim T\eta^{-1}g \qquad (16)$$

in order to satisfy the required condition for the true relaxation time, $\tau_{tr}$,

$$\tau_{tr} \sim \tau N^{1/2}. \qquad (17)$$

Here, $g \sim a^{-2} r^{-1}$ is a pair correlation function [13] (see also (4)) and we state that the torque has always to be applied with a certain (lever) length $r \sim aN^{1/2}$. In more realistic case $r \sim aN^{1-\nu}$ [13] where $\nu \approx 1 - 6^{-1/2} \approx 0.59$ [13] is the correlation radius exponent of swelling chain [15], since screening the volume interactions vanishes if the mechanical field is applied to a polymer melt. Consequently, the exponent sought $\approx 3.41$. Thus, the maximum time of relaxation (17) increases resulting in the hydrodynamic effect of velocity of a ball-shaped body, rotating effectively in a high-molecular-weight liquid with the viscosity $\eta$, on the friction force $f$. A characteristic size of this body is $r$, the coefficient of rotation diffusion $D_{rot}$ can be found from (16).

It is clear that the fluctuation attraction between all monomers of chain results in an inhibition of large-scale motion of chain ends as an average effect.

2.5 Ring macromolecules

It is impossible to compare the recent Monte Carlo research on linear and ring chains [18, 19] with our results for 2 reasons, although the investigations of ring macromolecules are continued expansively as seen from the literature [20-23]. The first one is reported by the authors. They are not capable of considering the case $N >> N_e$. The second one is the absence of hard criteria for decoupling the rotation and translation diffusion in their model for linear chains.

The diffusion coefficient for a melt of linear molecules can be estimated as follows:

$$D = R_0^2/\tau_{tr} \sim N^{(2\nu-1/2)2}/N^{4-\nu} \sim N^{-5(1-\nu)} \sim N^{-2.05}$$

where we have assumed $R_0 \sim R_g$ and have used the expression for the gyration radius $R_g \sim N^{2\nu-1/2}$ obtained by the author [13]. The reptation law $D \sim N^{-2}$ [15] can be expected for unknotted rings since there is no fluctuation attraction of the chain ends.

If $N < N_e$ then the formula (10) is true, and $D \sim (aN)^2/\tau_1 N^2 \sim D_1$ ($R_0 \sim aN$). Thus, for $N \sim N_e$ one can observe any exponent ranged from *-2.05* to *0*. In this respect let us recall the estimation $N_e \approx 283$ for linear chains [13].

2.6 Conclusion remarks

The rotation vibration motion can not be observed at macroscopic scales due to decreasing the field with increasing the distance. More accurate accordance with experimental data can be not achieved owing to the interaction potential depends on the chemical nature of chain monomers at short distances. An antisymmetric stress seems can appear as a macromolecule is capable of making rotation vibration precession motions. At large scales it transforms to an average tensor which will be symmetric probably.

We have also found that a non-equilibrium parameter of polymer melts such as the viscosity is indirectly connected with the equilibrium pair correlation function due to the finite length of macromolecules [13]. We can indeed neglect the fluctuation attraction of the chain ends, supposing it small, for long enough chains, i. e. if $N \to \infty$ and the physical state of the melt is not changed as in works [10, 11] then the exponent of viscosity relation to the molecular weight of polymer has to be *3*.

**Acknowledgements**

Thanks are expressed to Professor A.R.Khokhlov, to Professor I.Ya.Erukhimovich, and to Dr. V.Buchin from Moscow State University for useful discussions, to Professor S.N.Chvalun and to Professor A.N.Kraiko from Moscow Institute of Physics and Technology for helpful remarks, and also to Dr. A.V.Mironov for technical assistance in preparing the manuscript. The author is thankful to Russian Foundation for Basic Research (Grant No 01-03-32225) for financial support.

**Figure captions**

**Fig. 1.** A macromolecule crawling into entanglement network (open circles) can make the rotation vibration motion (shown by vectors) resulting in increasing viscosity. Dotted curve designates the initial position of the tube, and solid curve - its final position. A part of the initial contour coincides with the final one.

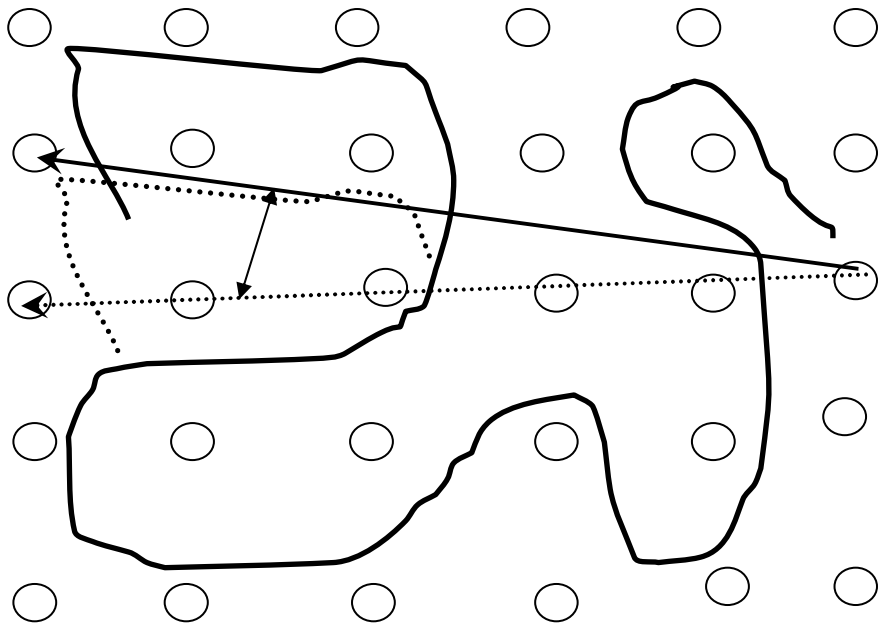